\documentclass[11pt,a4paper]{article}
\usepackage{float}
\usepackage{amsmath}
\usepackage{amsfonts}
\usepackage{amssymb}
\usepackage{graphicx}

\newtheorem{theorem}{Theorem}

\newtheorem{proposition}{Proposition}[section]

\newtheorem{remark}{Remark}

\newenvironment{proof}[1][Proof]{\noindent\textbf{#1.} }{\ \rule{0.5em}{0.5em}}

\let \a = \alpha
\let \b = \beta

\let \g = \gamma

\begin{document}

\title{On the similarity solutions for a steady MHD equation}

\author{Jean-David HOERNEL $\dag,^*$}
\date{}
\maketitle

\begin{center}
$\dag$ Department of Mathematics, Technion-Israel Institute of Technology\\
Amado Bld., Haifa, 32000 ISRAEL\\
E-mail: j-d.hoernel@wanadoo.fr
\end{center}

\abstract{In this paper, we investigate the similarity solutions for a steady laminar incompressible
 boundary layer equations governing the magnetohydrodynamic (MHD) flow near the forward stagnation point
  of two-dimensional and axisymmetric bodies. This leads to the study of a boundary value problem involving
  a third order autonomous ordinary differential equation. Our main results are the existence, uniqueness and
  nonexistence for concave or convex solutions.}

\footnotetext{\hspace{-0.8cm} MSC: 34B15, 34C11, 76D10}
\footnotetext{\hspace{-0.8cm} PACS: 47.35.Tv, 47.65.–d, 47.15.Cb}
 \footnotetext{\hspace{-0.8cm} Key words and
phrases: Boundary layer, similarity solution, third order
nonlinear differential equation,  boundary value problem, MHD. }

\footnotetext{\hspace{-0.8cm} $^*$ The author thanks the
Department of Mathematics of the Technion for supporting his
researches through a Postdoctoral Fellowship in the frame of the RTN ``Fronts-Singularities".}

\section{Introduction}
Boundary layer flow of an electrically
conducting fluid over moving surfaces emerges in a large variety of
industrial and technological applications. It has been investigated
by many researchers, Wu \cite{wu} has studied the effects of suction
or injection on a steady two-dimensional MHD boundary layer flow on
a flat plate, Takhar et al. \cite{thakhar1} studied a MHD asymmetric
flow over a semi-infinite moving surface and numerically obtained
the solutions. An analysis of heat and mass transfer characteristics
in an electrically conducting fluid over a linearly stretching sheet
with variable wall temperature was investigated by Vajravelu and
Rollins \cite{vajra}. In \cite{muha} Muhapatra and Gupta  treated the
steady two-dimensional stagnation-point flow of an incompressible
viscous electrically conducting fluid towards a stretching surface,
the flow being permeated by a uniform transverse magnetic field. For
more details see also
\cite{chakra-gupta}, \cite{kumari1}, \cite{pop}, \cite{takhar2}
and the references therein.\\
Motivated by the above works, we aim here to give analytical results
about the third order non-linear autonomous differential equation
\begin{equation}\label{equation}
    f'''+\frac{m+1}{2}ff''+m(1-{f'}^{2})+M(1-f')= 0 \qquad
    \mbox{on} \quad [0,\infty)
\end{equation}
 accompanied by the boundary conditions
\begin{equation}\label{cond03}
f(0)=a, \quad f'(0)=b, \quad  f'(\infty)=1
\end{equation}
 where $a, b, m, M \in \mathbb{R}$ and
$f'(\infty):=\underset{t\rightarrow\infty}{\lim} f'(t)$.
 Equation (\ref{equation}) is very interesting because it contains many
 known equations as particular cases. Let us give some examples.\\
 Setting $M=0$ in (\ref{equation}), leads to the well-known Falkner-Skan equation
 (see \cite{falkner},\cite{coppel},\cite{fdr} and the references
 therein). While the case $M=-m$ reduces (\ref{equation}) to equation that arises when considering the mixed
 convection in a fluid saturated porous medium near a semi-infinite vertical flat plate with prescribed
 temperature studied by many authors, we refer the reader to
 \cite{aly},\cite{brighi1},\cite{guedda},\cite{kumari2} and the
references therein. The case $M=m=0$
 is refereed  to the Blasius equation introduced in
 \cite{blasius} and studied by several authors (see for example
 \cite{brighi2},\cite{brighi3},\cite{utz}). Recently, the case
 $m=-1$ have been studied in \cite{aah} the authors show existence
 of "pseudo-similarity" solution, provided that the plate is permeable with suction. Mention may be made also to \cite{ghh}, where the authors show existence of an infinite number of similarity solutions for the case of a non-Newtonian fluid.\\
 More recently, some results have been obtained by Brighi and Hoernel \cite{brighi4}, about the more
 general equation
\begin{equation}\label{eq1.3}
    f'''+ff''+g(f')= 0 \qquad
    \mbox{on} \quad [0,\infty)
\end{equation}
with the boundary conditions
 \begin{equation}\label{eq1.4}
 f(0)=\alpha, \quad f'(0)=\beta, \quad  f'(\infty)=\lambda
 \end{equation}
 where $\alpha, \beta, \lambda \in \mathbb{R}$ and $g$ is a given
 function. Guided by the analysis of \cite{brighi4} we shall prove
 that problem (\ref{equation})-(\ref{cond03}) admits a unique concave or
 a unique convex solution for $m>-1$ according to the values of
 $M$.  We give also non-existence results for $m\in\mathbb{R}$ and
 related values of $M$.\\
\section{Flow analysis}
Let us suppose that an electrically conducting fluid (with
electrical conductivity $\sigma$) in the presence of a transverse
magnetic field $B(x)$ is flowing past a flat plate stretched with a
power-law velocity. According to \cite{aah},\cite{rao},\cite{sher},
such phenomenon is described by the following equations
\begin{equation}\label{eq1}
\frac{\partial{u}}{\partial{x}}+\frac{\partial{v}}{\partial{y}}=0,
\end{equation}
\begin{equation}\label{eq2}
u\frac{\partial{u}}{\partial{x}}+v\frac{\partial{u}}{\partial{y}}=u_e{u_e}_{x}+
\nu\frac{\partial^2{u}}{\partial{y^2}}+\frac{\sigma
B^2(x)}{\rho}(u_e-u).
\end{equation}
Here, the induced magnetic field is neglected. In a cartesian system
of co-ordinates $(O,x,y)$, the variables $u$ and $v$ are the
velocity components in the $x$ and $y$ directions respectively. We will denote by
$u_e(x) = \g x^m, \g>0$ the external velocity,
$B(x)=B_{0}x^{\frac{m-1}{2}}$  the applied magnetic field, $m$
the power-law velocity exponent, $\rho $ the fluid density and $
\nu$ the kinematic viscosity. \\
The boundary conditions for problem (\ref{eq1})-(\ref{eq2}) are
\begin{equation}\label{eq3}
u(x,0)=u_{w}(x)=\a x^{m},\qquad
v(x,0)=v_{w}(x)=\b x^{\frac{m-1}{2}},\qquad u(x,\infty)=u_e(x)
\end{equation}
where $u_{w}(x)$ and $v_{w}(x)$ are the stretching and the suction
(or injection) velocity respectively and $\a, \b $ are constants. Recall that $\a>0$ is referred to the suction,
$\a<0$ for the injection and $\a=0$ for the impermeable plate.\\
A little inspection shows that equations (\ref{eq1}) and (\ref{eq2})
accompanied by conditions (\ref{eq3}) admit a similarity solution.
Therefore, we introduce the dimensional stream function $\psi$ in
the usual way to get  the following equation
\begin{equation}\label{eq4}
\frac{\partial \psi}{\partial
{y}}\frac{\partial^{2}\psi}{\partial{x}\partial{y}}-\frac{\partial{\psi}}
{\partial{x}}\frac{\partial^{2}\psi}{\partial{y^2}}=u_e{u_e}_{x}+
\nu\frac{\partial^{3}\psi}{\partial{y^3}}+\frac{\sigma
B^2(x)}{\rho}(u_e -u).
\end {equation}
The boundary conditions become
\begin{equation}\label{eq5}
 \frac{\partial\psi}{\partial{y}}(x,0)=\a x^m,\,\
 \frac{\partial\psi}{\partial{x}}(x,0)=-\b x^\frac{m-1}{2},\,\
\frac{\partial\psi}{\partial{y}}(x,\infty)=\g x^m.
\end {equation}
Defining the similarity variables as follows
$$\displaystyle
\psi(x,y)= x^{\frac{m+1}{2}} f(t)\sqrt{\nu \g} \qquad
\mbox{and}\qquad t= x^{\frac{m-1}{2}}y\sqrt{\frac{\nu}{\g}}$$ and
substituting in equations (\ref{eq4}) and (\ref{eq5}) we get the
following boundary value problem
 \begin{equation}\label{eq6}
\left\{\begin{array}{l}
    f'''+\frac{m+1}{2}ff''+m(1-{f'}^{2})+M(1-f')= 0,\\
\\
    f(0)=a, \quad f'(0)=b, \quad  f'(\infty)=1
\end{array}\right.
\end{equation}
where $a=\frac{2\b}{(m+1)\sqrt{\nu \g}},\ b=\frac{\a}{\g}$ and
$ M=\frac{\sigma B_{0}^{2}}{\g \rho}>0$ \;is the Hartmann number and
the prime is for differentiating  with respect to $t$.\\
\section{Various results}
First, we give the following
\begin{remark}{\rm Let $b=1$, then the function $f(t)=t+a$ is a solution of the problem
$(\ref{equation})$-$(\ref{cond03})$ for any values of $m$ and $M$ in
$\mathbb{R}$. We cannot say much about the uniqueness of the
previous solution, but if $g$ is another solution with
$g''(0)=\gamma>0$ then, since $g'(0)=g'(\infty)=1$ there exists
$t_0>0$ such that $g'(t_0)>1$, $g''(t_0)=0$ and $g'''(t_0)\leq 0$.
However, from (\ref{equation}) we obtain that for $m>0$ and $M>0$,
$g'''(t_0)=-m(1-g'^2(t_0))-M(1-g'(t_0))>0$ and thus a
contradiction.}
\end{remark}
\bigskip
Suppose now that $f$ verifies the equation $(\ref{equation})$ only.
We will now establish some estimations for the possible extremals of
$f'$.
\begin{proposition} \label{pmin} Let $f$ be a solution of the equation $(\ref{equation})$ and $t_0$ be
a minimum for $f'$  (i.e. $f''(t_0)=0$ and $f'''(t_0)\geq 0$), if
it exists. For such a point $t_0$ we have the following
possibilities, according to the values of $m$ and $M$.

\noindent \begin{tabular}{ll} 
$\bullet$ For $m<0$ & -- if $M<-2m$, then $-1-\frac{M}{m}\leq f'(t_0) \leq 1$,\\
& -- if $M=-2m$, then $f'(t_0)=1$, \\
& -- if $M>-2m$, then $1\leq f'(t_0) \leq -1-\frac{M}{m}$.
\end{tabular}

\medskip
\noindent \begin{tabular}{ll}

$\bullet$ For $m=0$ & -- if $M<0$, then $f'(t_0) \leq 1$,\\
& -- if $M>0$, then $1\leq f'(t_0)$.
\end{tabular}
\medskip

\noindent \begin{tabular}{ll} 

$\bullet$ For $m>0$ & -- if $M<-2m$, then $f'(t_0) \leq 1$ or $-1-\frac{M}{m}\leq f'(t_0)$,\\
& -- if $M>-2m$, then $1\leq f'(t_0)$ or $f'(t_0)\leq-1-\frac{M}{m}$.
\end{tabular}
\end{proposition}
\begin{proof}
Let $t_0$ be a minimum of $f'$ with $f$ a solution of $(\ref{equation})$. Using the equation
(\ref{equation}) and the fact that $f''(t_0)=0$, we obtain that
$$f'''(t_0)+m(1-f'^2(t_0))+M(1-f'(t_0))=0.$$
Setting $p(x)=m(1-x^2)+M(1-x)$, we have that $f'''(t_0)\geq 0$ leads to $g(f'(t_0))\leq 0$ and the results follows.
 Let us remark that in both cases $m=M=0$ and $m>0$, $M=-2m$ we cannot deduce anything about $f'(t_0)$.
\end{proof}
\begin{proposition} \label{pmax} Let $f$ be a solution of the equation $(\ref{equation})$
and $t_0$ be a maximum for $f'$ (i.e. $f''(t_0)=0$ and
$f'''(t_0)\leq )$, if it exists. For such a point $t_0$ we have
the following possibilities, according to the values of $m$ and $M$.

\noindent \begin{tabular}{ll} 

$\bullet$ For $m<0$  & -- if $M<-2m$,
then $f'(t_0)\leq -1-\frac{M}{m}$
or $f'(t_0) \geq 1$,\\
& -- if $M>-2m$, then $f'(t_0) \leq 1$ or $f'(t_0)\geq -1-\frac{M}{m}$.
\end{tabular}
\medskip

\noindent \begin{tabular}{ll} $\bullet$ For $m=0$ & --  if $M<0$, then $f'(t_0) \geq 1$,\\
& -- if $M>0$, then $f'(t_0)\leq 1$.
\end{tabular}
\medskip

\noindent \begin{tabular}{ll} $\bullet$ For $m>0$ & -- if $M<-2m$, then $1\leq f'(t_0) \leq -1-\frac{M}{m}$,\\
& -- if $M=-2m$, then $f'(t_0)=1$,\\
& -- if $M>-2m$, then $-1-\frac{M}{m}\leq f'(t_0)\leq 1$.
\end{tabular}
\end{proposition}
\begin{proof}
We proceed as in the previous Proposition, but this time, with the
condition $g(f'(t_0))\geq 0$. Let us remark that in both of cases
$m<0$, $M=-2m$ and $m=M=0$ we cannot deduce anything about
$f'(t_0)$.
\end{proof}
\medskip

We will now use the two previous Propositions to deduce results
about the possible extremals for $f'$ with $f$ a solution of the
problem $(\ref{equation})$-$(\ref{cond03})$.
\begin{theorem}  Let $f$ be a solution of the problem $(\ref{equation})$-$(\ref{cond03})$, $t_0$ be
a minimum for $f'$ $($i.e. $f''(t_0)=0$ and $f'''(t_0)\geq 0$$)$, if
it exists, and $t_1$ be a maximum for $f'$ $($i.e. $f''(t_1)=0$ and
$f'''(t_1)\leq 0$$)$, if it exists. For such points $t_0$ and $t_1$,
we have the following possibilities for the values of $f'$.

\noindent \begin{tabular}{ll} $\bullet$ For $m<0$ & -- if $M<-2m$, then $-1-\frac{M}{m}\leq f'(t_0) \leq 1\leq f'(t_1)$,\\
& -- if $M=-2m$, then $f'(t_0)=1$, \\
& -- if $M>-2m$, then $1\leq f'(t_0) \leq -1-\frac{M}{m}\leq f'(t_1)$.
\end{tabular}
\medskip

\noindent \begin{tabular}{ll} $\bullet$ For $m=0$ & -- if $M<0$, then $f'(t_0) \leq 1\leq f'(t_1)$,\\
& -- if $M>0$, then $f'$ cannot vanish.
\end{tabular}
\medskip

\noindent \begin{tabular}{ll} $\bullet$ For $m>0$ & -- if $M<-2m$, then $f'(t_0) \leq 1\leq f'(t_1)\leq -1-\frac{M}{m}$,\\
& -- if $M=-2m$, then $f'(t_1)=1$,\\
& -- if $M>-2m$, then $f'(t_0)\leq -1-\frac{M}{m}\leq f'(t_1)\leq 1$.
\end{tabular}
\end{theorem}
\begin{proof}
Taking into account the fact that $f'\to 1$ for large $t$ and
combining Proposition $\ref{pmin}$ and Proposition $\ref{pmax}$ lead
to the results.
\end{proof}
\begin{remark}
{\rm A consequence of the previous Theorem is that, for $m=0$ and
$M>0$ all the solutions of the problem
$(\ref{equation})$-$(\ref{cond03})$ have to be concave or convex
everywhere.}
\end{remark}
\section{The concave and convex solutions}
In this section we will first prove that, under some hypotheses, the
problem (\ref{equation})-(\ref{cond03}) admits a unique concave
solution or a unique convex solution for $m>-1$. Then, we will give
some nonexistence results about the concave or convex solutions for
$m\in\mathbb{R}$ according to the values of $M$. To this aim, we
will use the fact that, if $f$ is a solution of the problem
(\ref{equation})-(\ref{cond03}), then the function $h$ defined by
\begin{equation}
f(t)= \sqrt{\frac{2}{m+1}}\ h\left(\sqrt{\frac{m+1}{2}}t
\right) \label{h}
\end{equation}
with $m>-1$, is a solution of the equation
\begin{equation}
h'''+hh''+g(h')=0 \label{e1}
\end{equation}
on $[0,\infty)$, with the boundary conditions
\begin{equation}
h(0)=\sqrt{\frac{m+1}{2}}\ a, \quad h'(0)=b, \quad h'(\infty)=1\label{c1}
\end{equation}
and where
\begin{equation}
g(x)=\frac{2m}{m+1}(1-x^2)+\frac{2M}{m+1}(1-x). \label{g}
\end{equation}
In the remainder of this section we will made intensive use of the
results found in the paper \cite{brighi4} by Brighi and Hoernel.
\begin{remark}{\rm
It is immediate that for any $a\in \mathbb{R}$, if $b<1$ there is no
concave solutions of the problem $(\ref{equation})$-$(\ref{cond03})$
and if $b>1$  there is no convex solutions of the problem
$(\ref{equation})$-$(\ref{cond03})$.}
\end{remark}
\subsection{Concave solutions}
Let us begin with the two following results about existence,
uniqueness and nonexistence of concave or convex solutions for the
problem $(\ref{equation})$-$(\ref{cond03})$.
\begin{theorem}\label{concave}
Let $a\in \mathbb{R}$ and $b>1$. Then, there exists a unique concave
solution of the problem $(\ref{equation})$-$(\ref{cond03})$ in the
two following cases 

\noindent \begin{tabular}{l}
$\bullet$ $-1<m\leq 0$ and $M>-m(b+1)$,\\
$\bullet$ $m>0$ and $M\geq-2m$.
\end{tabular}

\noindent Moreover, there exists $a<l<\sqrt{a^2+4\frac{b-1}{m+1}}$ such that
$\underset{t\to \infty}{\lim}\{f(t)-(t+l)\}=0$ and for all $t\geq
0$ we have $t+a\leq f(t)\leq t+l$.
\end{theorem}
\begin{proof}
Let $f$ be a solution of the problem $(\ref{equation})$-$(\ref{cond03})$ with $m>-1$ and consider the
 function $h$ that is defined by (\ref{h}) and that verifies $(\ref{e1})$-$(\ref{c1})$.
Then, as $g(1)=0$ for the function $g$ defined by (\ref{g}), using Theorem 1 of \cite{brighi4} we get
that the problem (\ref{e1})-(\ref{c1}) admits a unique concave solution $h$ for
every $a\in \mathbb{R}$ and $b>1$ if and only if $g(x)<0$ for all
$x$ in $(1,b]$. Noticing that $g\left(-\frac{M}{m}-1\right)=0$, the previous condition is verified
if $m\geq 0$ and $-\frac{M}{m}-1\leq1$, if $m=0$ and $M>0$ or if
$-1<m<0$ and $M>-2m$ and $b<-\frac{M}{m}-1$.
Using now the Proposition 1 of \cite{brighi4} for $h$ we have the second
result.
\end{proof}
\begin{theorem}\label{noconcave}
Let $b>1$. Then, there are no concave solutions of the problem
$(\ref{equation})$-$(\ref{cond03})$ in the following cases \noindent
\begin{tabular}{l}
$\bullet$ $a\in \mathbb{R}$, $m\leq -1$ and $M\leq -2m$,\\
$\bullet$ $a\leq 0$, $-1<m<-\frac{1}{3}$ and $M\leq -\frac{5m+1}{2}$,\\
$\bullet$ $a\leq 0$, $m=-\frac{1}{3}$ and $M<\frac{1}{3}$,\\
$\bullet$ $a<0$, $m=-\frac{1}{3}$ and $M=\frac{1}{3}$,\\
$\bullet$ $a\leq 0$, $m>-\frac{1}{3}$ and $M\leq -\frac{(3m+1)b+2m}{2}$.
\end{tabular}
\end{theorem}
\begin{proof}
Let $a\in \mathbb{R}$, $m\leq -1$ and $f$ be a concave solution of the problem $(\ref{equation})$-$(\ref{cond03})$.
We then have that $f'>1$, $f''<0$, $f'''>0$ everywhere and $f(t)>0$ for $t$ large enough because $f'(t)\to 1$ as
$t\to \infty$. Using the fact that $\frac{m+1}{2}ff''>0$ near infinity, we obtain from (\ref{equation}) that
$$f'''\leq -m(1-f'^2)-M(1-f')$$
near infinity. As the polynomial function $-m(1-x^2)-M(1-x)$ is
negative for all $x$ in $[1,\infty]$ if $m\leq -1$ and $M\leq -2m$,
we get that $f'''<0$ near infinity because $f'>1$ everywhere. This
is a contradiction, so concave solutions cannot exist in this case.
Consider now $m>-1$ and $h$ a solution of the problem
$(\ref{e1})$-$(\ref{c1})$. Let us define the function $\hat g$
 by $\hat g(x)=g(x)-x^2+x$, a simple calculation leads to
$$\hat g(x)=\frac{1}{m+1}\left( -(3m+1)x^2+(m+1-2M)x+2(m+M)\right).$$
Then, the Theorem 2 from \cite{brighi4} tells us that problem $(\ref{e1})$-$(\ref{c1})$ admits no concave solutions
 for $a\leq 0$ if $\forall x\in[1,b]$, $\hat g(x)\geq 0$ and
$-a+\max_{x\in[1,b]}\ \hat g(x)>0$. These conditions lead to the  results for problem
$(\ref{equation})$-$(\ref{cond03})$ with $m>-1$.
\end{proof}

\medskip
The results from Theorem \ref{concave} and Theorem \ref{noconcave}
are summarized in the  Figure $1$ in which the
 plane $(m,M)$ contains three disjoints regions A, B and C are
 defined as  
 
\noindent -- A: Existence of a unique concave solution for $m>-1$, $b>1$ and $a\in\mathbb{R}$,\\
\noindent -- B: No concave solutions for $m>-1$, $b>1$ and $a\leq 0$,\\
\noindent -- C: No concave solutions for $m\leq -1$, $b>1$ and
$a\in\mathbb{R}$.
\newpage
$$\includegraphics[]{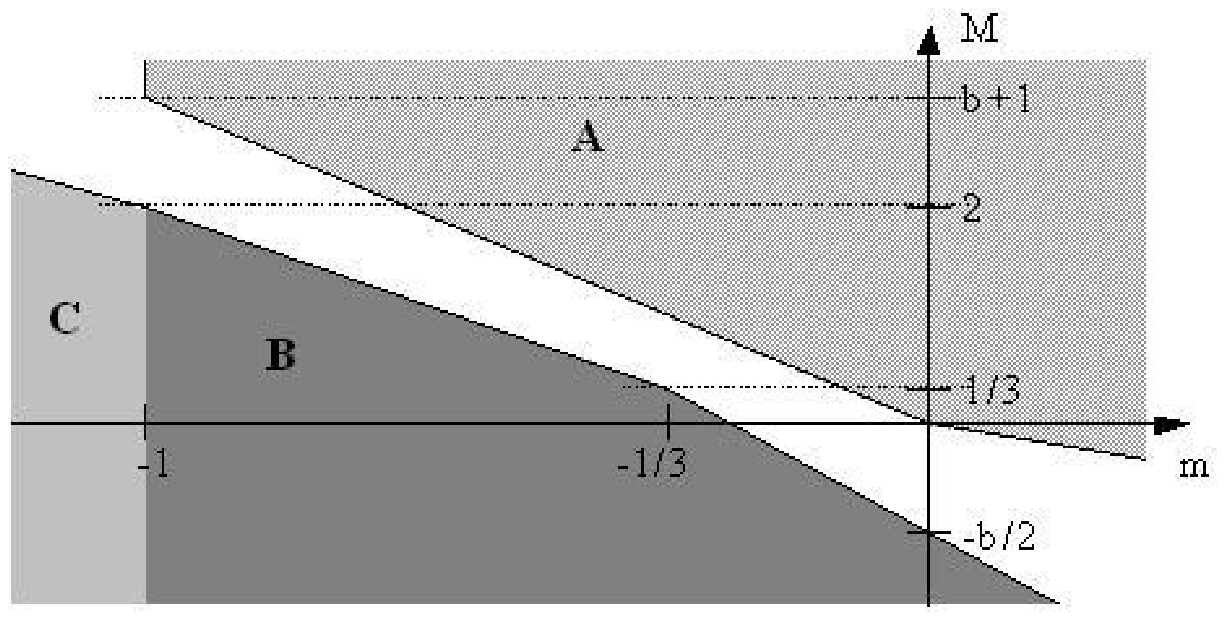}$$
$$\text{Figure }1$$
\subsection{Convex solutions}
We will now give existence, uniqueness and non-existence results for
the convex solutions of the problem
$(\ref{equation})$-$(\ref{cond03})$.
\begin{theorem}\label{convex}
Let $a\in \mathbb{R}$ and $0\leq b<1$. Then, there exists a unique
convex solution of the problem $(\ref{equation})$-$(\ref{cond03})$
in the following cases

\noindent \begin{tabular}{l}
$\bullet$  $-1<m<0$ and $M\geq -2m$,\\
$\bullet$ $m\geq 0$ and $M> -m(b+1)$.
\end{tabular}

\noindent Moreover, there exists $l>a$ such that $\underset{t\to
\infty}{\lim}\{f(t)-(t+l)\}=0$ and for all $t\geq 0$ we have
$t+a\leq f(t)\leq t+l$.
\end{theorem}
\begin{proof}
We proceed the same way as for Theorem \ref{concave}, but with the
condition that $g(x)>0$ for all $x$ in $[b,1)$. We conclude by using first the Theorem
3 from \cite{brighi4}, then the Proposition 2 from \cite{brighi4}.
\end{proof}
\medskip
\begin{theorem}\label{noconvex}
Let $0\leq b<1$. Then, there are no convex solutions of the problem
$(\ref{equation})$-$(\ref{cond03})$ in the following cases 

\noindent
\begin{tabular}{l}
$\bullet$ $a\in \mathbb{R}$, $m\leq -1$ and $M\leq -m(b+1)$,\\
$\bullet$ $a\leq 0$, $-1<m<-\frac{1}{3}$ and $M\leq -\frac{(3m+1)b+2m}{2}$,\\
$\bullet$ $a\leq 0$, $m=-\frac{1}{3}$ and $M<\frac{1}{3}$,\\
$\bullet$ $a<0$, $m=-\frac{1}{3}$ and $M=\frac{1}{3}$,\\
$\bullet$ $a\leq 0$, $m>-\frac{1}{3}$ and $M\leq -\frac{5m+1}{2}$.
\end{tabular}
\end{theorem}
\begin{proof}
For $m>-1$ and $a\leq 0$, the proof is the same as the previous one,
but this time we need that $\forall x\in[b,1]$, $\hat g(x)\leq 0$
and $-a+\max_{x\in[b,1]}\ \hat g(x)>0$, according to the Theorem 4
from \cite{brighi4}. Consider now $m\leq -1$, $a\in\mathbb{R}$ and
let $f$ be a convex solution of the problem
$(\ref{equation})$-$(\ref{cond03})$. We have that $b\leq f'<1$,
$f''>0$, $f'''<0$ everywhere and that $f(t)>0$ for $t$ large enough
because $f'(t)\to 1$ as $t\to \infty$. According to equation
(\ref{equation}), we have that
$$f'''=-\frac{m+1}{2}ff''-m(1-f'^2)-M(1-f')$$
with $-\frac{m+1}{2}ff''>0$ near infinity.  As the polynomial function $-m(1-x^2)-M(1-x)$ is
 positive for all $x$ in $[b,1]$ if $m\leq -1$ and $M\leq -m(b+1)$, we get that $f'''>0$ near
 infinity because $b\leq f'<1$. This is a contradiction, thus convex solutions cannot exist in this case.
\end{proof}
\medskip

The results from Theorem \ref{convex} and Theorem \ref{noconvex} are
summarized in the Figure. 2 in which the plane $(m,M)$ contains
three disjoints regions A, B and C that corresponds to 

\noindent --
A: Existence of an unique convex solution for $m>-1$, $0\leq b<1$
and $a\in\mathbb{R}$, 

\noindent -- B: No convex solutions for
$m>-1$, $0\leq b<1$ and $a\leq 0$,
 
 \noindent -- C: No convex
solutions for $m\leq -1$, $0\leq b<1$ and $a\in\mathbb{R}$.
$$\includegraphics[]{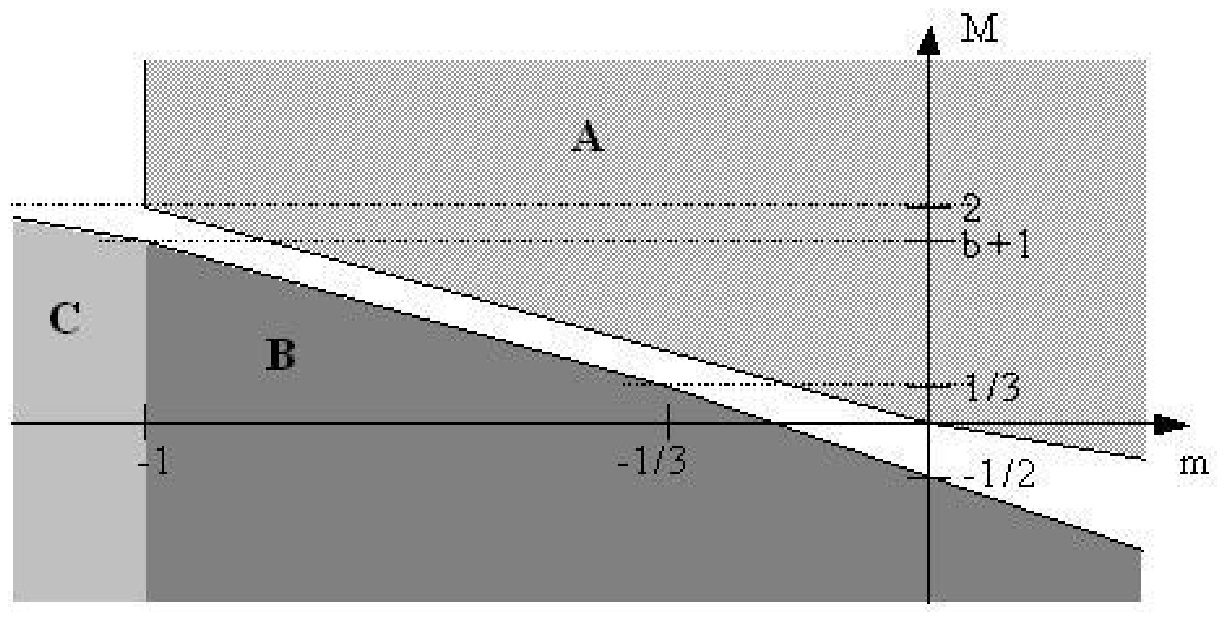}$$
$$\text{Figure }2$$
\section{Conclusion}
In this paper, we have shown the existence of a unique concave or a
unique convex solution of the problem
$(\ref{equation})$-$(\ref{cond03})$ for $m>-1$, according to the
values of $M$. We also have obtained nonexistence results for $m\in
\mathbb{R}$ and related values of $M$, as well as some clues about
the possible behavior of $f'$. This paper is a first work on this
problem, there is still much left to do because of its complexity.
Notice that the case $M=-2m$ plays a particular  role in the problem
$(\ref{equation})$-$(\ref{cond03})$, because it is the only one for
which we are able to predict the possible changes of concavity for
$f$. Its study will be the subject of a forthcoming paper.

\section*{Acknowledgement}
The author would like to thank Prof. B. Brighi for his many advices
and for introducing him to the similarity solutions family of
problems.

\end{document}